\documentclass{aa}  

\usepackage{graphicx,amsmath,amssymb}
\usepackage{orcidlink}
\usepackage{txfonts,xcolor}
\newcommand{\chang}[1]{#1}

\begin{document}

   \title{A measurement of the escaping ionising efficiency of galaxies \\ at redshift $5$}

   \author{S.~E.~I.~Bosman
          \inst{1, 2}\orcidlink{0000-0001-8582-7012}
          \and
          F.~B.~Davies\inst{2}\orcidlink{0000-0003-0821-3644}\fnmsep
          }

   \institute{Institute for Theoretical Physics, Heidelberg University, Philosophenweg 12, D–69120, Heidelberg, Germany\\ \email{bosman@thphys.uni-heidelberg.de}
         \and
             Max-Planck-Institut für Astronomie, Königstuhl 17, 69117 Heidelberg, Germany
             }

   \date{Received XXX; accepted XXX}

 
  \abstract
   {The escaping ionising efficiency from galaxies, $f_{\rm esc}\xi_{\rm ion}$, is a crucial ingredient for understanding their contribution to hydrogen reionisation, but both of its components, $f_{\rm{esc}}$  and $\xi_{\rm{ion}}$, are extremely difficult to measure.
  We measure the average escaping ionising efficiency $\langle f_{\rm{esc}} \xi_{\rm{ion}}\rangle$ of galaxies at $z=5$ implied by the mean level of ionisation in the intergalactic medium via the Lyman-$\alpha$ forest.
We use the fact that $\dot{N}_{\rm{ion}} = \rho_{\rm{UV}} f_{\rm{esc}} \xi_{\rm{ion}}$, the product of the ionising output and the UV density $\rho_{\rm{UV}}$, can be calculated from the known average strength of the UV background and the mean free path of ionising photons. These quantities, as well as $\rho_{\rm{UV}}$, are robustly measured at $z\leq6$. We calculate the missing factor of $\langle f_{\rm{esc}} \xi_{\rm{ion}}\rangle$ at $z=5$, during a convenient epoch after hydrogen reionisation has completed and the intergalactic medium has reached ionisation equilibrium, but before bright quasars begin to dominate the ionising photon production. 
Intuitively, our constraint corresponds to the required escaping ionising production from galaxies in order to avoid over- or under-ionising the Lyman-$\alpha$ forest.
   We obtain a measurement of $\log \langle f_{\rm{esc}} \xi_{\rm{ion}}\rangle /$erg Hz$^{-1}$ $ = 24.28_{-0.20}^{+0.21}$  at $z=5$ when integrating $\rho_\text{UV}$ down to a limiting magnitude $M_\text{lim}=-11$.
  Our measurement of the escaping ionising efficiency of galaxies is in rough agreement with both observations and \chang{most} models.}


   \maketitle
%

\section{Introduction}

The process of reionisation is thought to be driven by galaxies at $5.5\lesssim z \lesssim 10$. Recently, the discoveries of unexpectedly high numbers of galaxies at $z>10$ \citep{Finkelstein24} and of their ability to carve out large ionised regions at $z\gtrsim7$ \citep{MG20,TT24} have raised concerns that early galaxies would complete reionisation too early compared to constraints from the Lyman-$\alpha$ forest (\citealt{Munoz24,Simmonds24}; but see \citealt{DBF24}).  

To reconcile observations of early galaxies, renewed interest has arisen for precise indirect diagnostics of the production rate of ionising photons by early galaxies (e.g.~\citealt{Tang19}) and the escape fraction of these photons from their hosts (e.g.~\citealt{Jaskot24a,Jaskot24b}). In this paper, we point out that the product of these quantities -- the escaping ionising output of galaxies per unit UV production -- is already known fairly precisely at $z=5$. Our constraints provide an independent consistency check of galaxy evolutionary models and indirect tracers of ionisation production right after the end of reionisation.

\section{Methods}

A popular way to conceptualise the progress of the hydrogen reionisation is via the Madau equation \citep{Madau99}:
\begin{equation}\label{m99}
\frac{dQ_\text{HII}}{dt} = \frac{\dot{N}_\text{ion}}{\langle n_\text{H} \rangle} - \frac{Q_\text{HII}}{t_\text{rec}},
\end{equation}
where the effects of the production of ionising photons from galaxies at a rate $\dot{N}_\text{ion}$ and the recombination of ionised hydrogen atoms inside of sinks on a characteristic timescale $t_\text{rec}$ combine to increase the ionised fraction of hydrogen $Q_\text{HII}$. 

\begin{figure*}
    \centering
    \includegraphics[width=\textwidth]{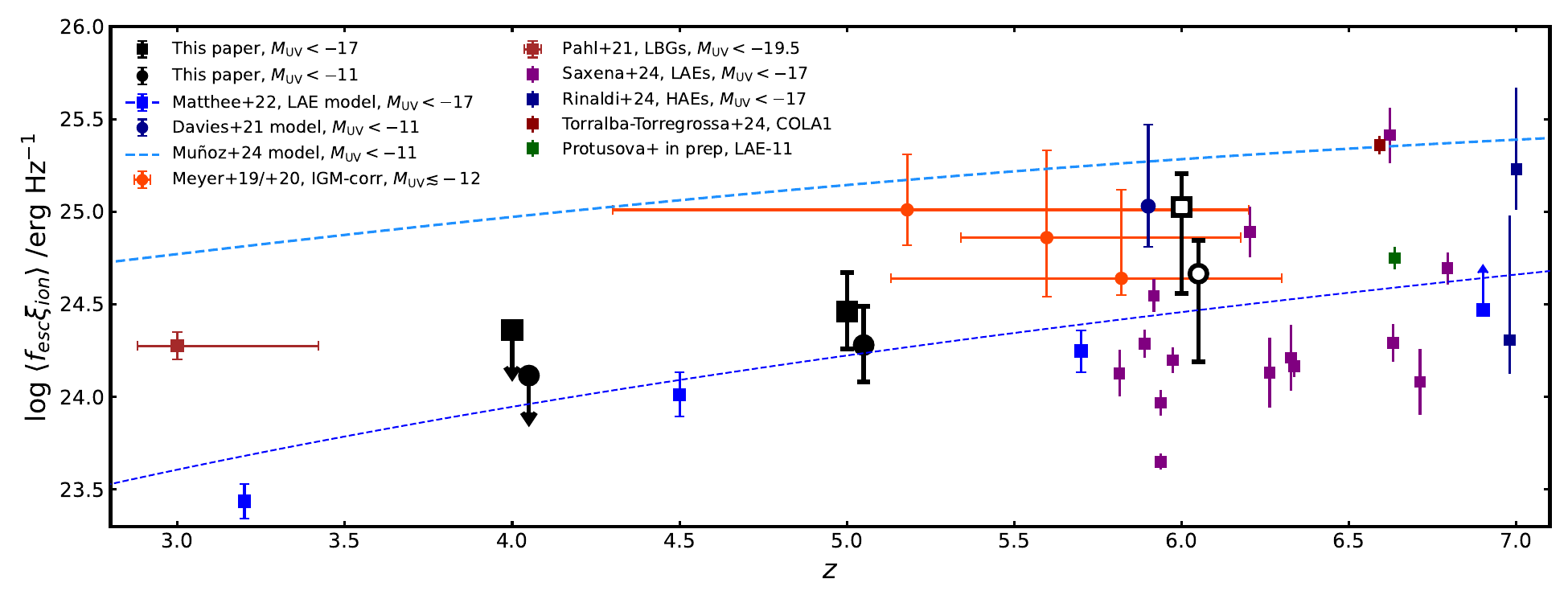}
    \vskip-1em
    \caption{Measurements of $\langle f_\text{esc}\xi_\text{ion}\rangle$ across redshift. Our constraints based on $\rho_\text{UV}$ and $\dot{N}_\text{ion}$ are shown for two different $M_\text{lim}$ (black). For comparison, we show measurements from the galaxy-optical depth cross-correlation \citep{Meyer19,Meyer20}, direct detections of the Lyman continuum \citep{Pahl21}, and indirect $f_\text{esc}$ and $\xi_\text{ion}$ tracers \citep{Saxena24,Rinaldi24,TT24}. The implications of the LAE-driven reionisation model of \citet{Matthee22}, as well as the models of \citet{Munoz24} and \citet{Davies21} are shown in shades of blue. At $z=4$, our argument only produces an upper limit due to the unknown contribution of quasars to $\dot{N}_\text{ion}$. At $z=6$, the measurement of $\dot{N}_\text{ion}$ becomes strongly dependent on the modeling of hydrogen reionisation, and our constraints should only be taken as a rough estimate.}
    \label{fig1}
\end{figure*}

Both terms in this equation are elusive and the focus of much theoretical and observational work. Here we focus on the first, ``source'', term, which is customarily written in terms of the bulk properties of the ionising galaxies:
\begin{equation}\label{eq2}
    \dot{N}_\text{ion} = \rho_\text{UV} f_\text{esc} \xi_\text{ion},
\end{equation}
where $\rho_\text{UV}$ is the UV luminosity density, $\xi_\text{ion}$ is the production rate of ionising photons ($\lambda<912$\AA) per unit UV production, and $f_\text{esc}$ is the escape fraction of ionising photons from galaxies. In particular, we will constrain the product of the latter two terms, $f_\text{esc}\xi_\text{ion}$ -- the escaping ionising efficiency.

The most interesting regime for $\langle f_\text{esc} \xi_\text{ion}\rangle$ is the epoch of reionisation itself, at $z\geq5.3$ \citep{Bosman22}. However, any measurements at later times are still valuable to constrain models of galaxy formation during the epoch of reionisation. At $z=5$, before \chang{bright} quasars begin to contribute significantly to $\dot{N}_\text{ion}$ (e.g.~\citealt{FB22}) but after the establishment of a homogeneous UV background in the intergalactic medium (IGM), both $\rho_\text{UV}$ and $\dot{N}_\text{ion}$ have been measured fairly robustly.

\subsection{$\dot{N}_\text{ion}$}

The global production rate of ionising photons $\dot{N}_\text{ion}$ is closely related to $\Gamma$, the average ionisation rate in the IGM (e.g.~\citealt{Gaikwad23,Davies24}). Below, we briefly reproduce the mathematical argument linking the two parameters.

Under the ``local source'' approximation \citep{MW03}, the intensity of ionising radiation in the IGM is roughly equal to the product of the ionising emissivity $\epsilon(\nu)$ and the mean free path of ionising photons $\lambda(\nu)$: 
\begin{equation}
J(\nu) = \epsilon(\nu) \lambda(\nu) / 4\pi.
\end{equation}

We can then re-write $\dot{N}_{\rm ion}$ as an integral over the ionising emissivity, 
\begin{equation}\label{eq:4}
    \dot{N}_{\rm ion} = \int_{\nu=\nu_{\rm \chang{LyC}}}^{+\infty} \frac{\epsilon(\nu)}{h\nu} d\nu = \frac{\epsilon(\nu_{\rm LyC})}{h\alpha}, 
\end{equation}
where $\alpha$ encodes the frequency dependence of emission from sources past the Lyman-limit $\nu_\text{LyC}$:
\begin{equation}
    \epsilon(\nu) = \epsilon(\nu_{\text{LyC}}) \left(\nu/\nu_\text{LyC}\right)^{-\alpha}. 
\end{equation}
This leads to a direct link between $\Gamma$ and $\dot{N}_\text{ion}$:
\begin{eqnarray}\label{eq:main}
    \Gamma &=& \int_{\nu=\nu_\text{LyC}}^{+\infty} \frac{J(\nu)}{h\nu}\sigma_{\rm HI}(\nu)d\nu \nonumber \\ &=& \int_{\nu=\nu_\text{LyC}}^{+\infty} \frac{\dot{N}_\text{ion}\lambda(\nu)}{h^2 \alpha \nu} \left( \frac{\nu}{\nu_\text{LyC}} \right)^{-\alpha} \sigma_\text{HI}(\nu) d\nu,
\end{eqnarray}
where $\sigma_\text{HI}$ is the photoionisation cross-section of neutral hydrogen. It is frequently assumed that  $\lambda(\nu)=\lambda_\text{LyC}(\nu/\nu_\text{LyC})^{-0.9}$ \citep{Gaikwad23,Davies24}, where   $\lambda_\text{LyC}$ is the mean free path of photons at the Lyman limit. \chang{Instead of extending to infinity, the integrals in equations \ref{eq:4} and \ref{eq:main} are truncated at $4\nu_\text{LyC}$, where the spectra of galaxies have a sharp break at the helium-ionising edge.}

In practice, $\Gamma$ is measured from the mean transmitted flux in the Lyman-$\alpha$ forest by means of comparison with numerical simulations. While newer measurements have employed radiative-transfer simulations which track the history of reionisation, simpler simulations with a homogeneous UV background (UVB) are sufficient at $z<5.3$, where they give consistent results \citep{BB13,D'Aloisio18,Bosman22,Davies24}. 

Eq.~\ref{eq:main} also requires a measurement of $\lambda_\text{LyC}$. 
At $z=5$, the value of $\lambda_\text{LyC}$ measured via comparison with simulations by \citet{Gaikwad23} is in agreement with measurements from direct stacking of quasar spectra from \citet{Worseck14}, \citet{Becker21}, and \citet{Zhu23}. %

We adopt $\dot{N}_\text{ion} = 0.563^{+0.352}_{-0.207} \times10^{51}$ photons$/s /\text{Mpc}^3$ at $z=5$ computed in this manner by \citet{Gaikwad23}. The dominating source of uncertainty in the measurement of $\dot{N}_\text{ion}$ at $z=5$ is not the measurement precision of $\Gamma$ or $\lambda_\text{LyC}$, but the conversion in Eq.~\ref{eq:main} via the spectral parameter $\alpha$. A value of $\alpha=2$ is often assumed by default (e.g.~\citealt{BB13}), but if the sources are bluer on average past the Lyman-limit, then $\alpha$ and $\dot{N}_\text{ion}$ could be larger. \chang{In the above measurement from \citet{Gaikwad23}, a wide range $1.4\leq\alpha\leq2.6$ is marginalised over, and multiple other observational and model-dependent uncertainties are included. Interested readers will find a thorough quantitative discussion of the uncertainties involved in converting $\Gamma$ to $\dot{N}_\text{ion}$ in \citet{BB13}. Overall, we consider the uncertainties on the measurement of \citet{Gaikwad23} to be conservative.}


\subsection{$\rho_\text{UV}$}

The UV luminosity function ($\Phi(M_\text{UV})$, the UVLF) of galaxies at $z=5$ is well-constrained by observations from the \textit{Hubble} Space Telescope down to UV magnitudes $M_\text{UV}\sim-17$ \citep{Finkelstein15,Bouwens21} and down to $M_\text{UV}\sim-14$ using lensing fields \citep{Bouwens22}. The UV luminosity density, $\rho_\text{UV}$, can be calculated directly by integrating the UVLF down to a chosen limiting UV magnitude $M_\text{lim}$:
\begin{equation}
    \rho_\text{UV} = \int_{-\infty}^{M_\text{UV}=M_\text{lim}} \Phi(M_\text{UV}) dM_\text{UV}.
\end{equation}

Using the UVLF measurements of \citet{Bouwens21} and taking into account the uncertainties on the overall normalisation of the UVLF, we obtain $\rho_\text{UV}=1.93_{-0.32}^{+0.39}\times10^{26}$ erg$/\text{s}/\text{Mpc}^3/\text{Hz}$ for $M_\text{lim}=-17$, and $\rho_\text{UV}=2.96_{-0.49}^{+0.60}\times10^{26}$ erg$/\text{s}/\text{Mpc}^3/\text{Hz}$ for $M_\text{lim}=-11$, extrapolating below the detection limit of current telescopes. We note that the faint-end slope of the UVLF at $z=5$ is sufficiently shallow to allow $\rho_\text{UV}$ to converge when integrated to hypothetical infinitely faint galaxies; doing so only increases $\rho_\text{UV}$ by a further $5\%$ compared to assuming $M_\text{lim}=-11$. 

\chang{The integrated $\rho_\text{UV}$ is consistent across alternative determinations of the UVLF. Using the fit obtained by \citet{Bouwens22} results in values of $\rho_\text{UV}$ which are $2\%$ and $8\%$ higher than the nominal values above for $M_\text{lim}=-11, -17$ respectively. Using the results of \citet{FB22}, who parametrise the UVLF as a double-power-law with a turn-over instead of a Schechter function, results in $\rho_\text{UV}$ being $2\%$ and $10\%$ lower than the nominal values for $M_\text{lim}=-11, -17$ respectively. This variance is comfortably encompassed in the $1\sigma$ uncertainties of the inference based on \citet{Bouwens21}.}

\section{Results}

Inserting the measurements of $\rho_\text{UV}$ and $\dot{N}_\text{ion}$ into Eq.~\ref{eq2}, the only unknown quantity is the global $\langle f_\text{esc}\xi_\text{ion}\rangle$ of the galaxies contributing to $\rho_\text{UV}$. 
In the context of Eq.~\ref{m99}, the global $\langle f_\text{esc}\xi_\text{ion}\rangle$ can be thought of as the output of all galaxies sourcing the UVB; therefore it is sensible to extrapolate the UVLF to $M_\text{lim}=-11$. We obtain $\log \langle f_\text{esc}\xi_\text{ion}\rangle/$erg Hz$^{-1}$ $=24.28_{-0.20}^{+0.21}$ (Figure~\ref{fig1}). This value is slightly lower ($\sim 2\sigma$) than constraints from the cross-correlation of the Lyman-$\alpha$ forest with the location of galaxies \citep{Kakiichi18,Meyer19,Meyer20}, which attribute the detection of an IGM ionisation boost to faint galaxies clustered with the bright ones they detect.

It is interesting to compare our constraints with the model of \citet{Munoz24} where $\xi_\text{ion}$ and $f_\text{esc}$ increase sharply towards fainter $M_\text{UV}$, motivated by recent measurements with the \textit{James Webb} Space Telescope (JWST) of $\xi_\text{ion}$ \citep{Simmonds23} and low-redshift $f_\text{esc}$ scaling relations \citep{Chisholm22}. To do this, we reproduce their model and obtain the total $\dot{N}_\text{ion}$ at $z=5$ by integrating the ionising output of galaxies down to $M_\text{lim}=-11$. We then obtain the effective global $\langle f_\text{esc} \xi_\text{ion} \rangle$ by dividing $\dot{N}_\text{ion}$ by the corresponding $\rho_\text{UV}$ and plot the result in Figure~\ref{fig1} \chang{as a light blue dashed line}. 

The model of \citet{Munoz24} violates the global constraints from the Lyman-$\alpha$ forest by almost an order of magnitude\chang{.
In simple terms, our measurements come from dividing the number of ionising photons in the IGM (which is know from measurements of $\Gamma$) by the total UV emission of all galaxies in order to obtain the mean galaxy escaping emissivity (which is defined per unit UV emission). The galactic property scaling relations theorised by \citet{Munoz24} result in a mean escaping emissivity almost $10$ times higher, which is very strongly ruled out by our argument. W}e interpret this as signifying that not all assumptions in their model can be correct at once. \chang{An extensive discussion concerning the robustness of the various assumptions of their model can be found in \citet{Munoz24}.}

We repeat the calculation for $M_\text{lim}=-17$ and obtain $\log \langle f_\text{esc}\xi_\text{ion}\rangle/$erg Hz$^{-1}$ $=24.46_{-0.20}^{+0.21}$. This assumption corresponds to a scenario where only bright galaxies contribute to the UVB, with a complete cut-off at $M_\text{lim}$. A nice comparison can be made with the models of \citet{Matthee22}, who argue that bright Lyman-$\alpha$ emitters (LAEs) are sufficient to power reionisation. Our corresponding measurement is in slight tension ($1-1.5\sigma$) with their scenario.

For the interest of readers, we present constraints arising from the same argument at $z=4$ and $z=6$, but with the following strong caveats. At $z=4$, a large fraction of $\dot{N}_\text{ion}$ is attributable to luminous quasars whose own UVLF is increasing rapidly \citep{Kulkarni19qlf,FB22}. In Eq.~\ref{eq2}, $\dot{N}_\text{ion}$ should be reduced by the same fraction in order to back out the $\langle f_\text{esc}\xi_\text{ion}\rangle$ from galaxies alone. 
To avoid making a guess as to the quasar contribution to $\dot{N}_\text{ion}$, we use the $\rho_\text{UV}$ arising from the galaxy UVLF only \citep{Bouwens21}. Under this assumption and using the value of $\dot{N}_\text{ion}$ from \citet{BB13}, our measurement becomes a stringent upper limit of $\log \langle f_\text{esc}\xi_\text{ion}\rangle /$erg Hz$^{-1}$ $< 24.36$ at $z=4$; if this value is exceeded, then the galaxies alone would over-ionise the IGM. The upper limit is not in tension with the model of \citet{Matthee22}, but implies a tension with the $z\sim3$ direct measurements of \citet{Pahl21} which was also noted by those authors. The limit is also in strong tension with the model of \citet{Munoz24}.

\chang{The same caveat concerning an unknown contribution from active galactic nuclei could potentially apply to our main result at $z=5$ in light of recent studies suggesting a larger role of faint quasars in finishing hydrogen reionisation \citep{Grazian24,Madau24}. At the present time, it is not clear how this scenario can avoid over-heating the IGM \citep{D'Aloisio17} or why the same faint quasars would not initiate the reionisation of helium, which is known to only get underway at $z\lesssim4$ \citep{Gaikwad21}. A significant contribution of faint quasars would formally turn our measurements at $z=5$ into upper limits, exacerbating the tension with the model of \citet{Munoz24}; in that model, galaxies by themselves already over-ionise the Lyman-$\alpha$ forest by almost an order of magnitude.}

At $z=6$, we repeat the same process using the UVLF of \citet{Bouwens21} and the measurement of $\dot{N}_\text{ion}$ from $\Gamma$ of \citet{Gaikwad23}. 
The calculation of $\dot{N}_\text{ion}$ depends linearly on the measurement of $\lambda_\text{LyC}$, which differs significantly between current studies. To incorporate this uncertainty, we rescale the value of $\dot{N}_\text{ion}$ from \citet{Gaikwad23} to use the more direct measurement of $\lambda_\text{LyC}$ from \citet{Zhu23}, but extend the uncertainties to include the $-1\sigma$ limit of \citet{Gaikwad23}'s self-consistent measurement of $\lambda_\text{LyC}$. 
We obtain $\log \langle f_\text{esc}\xi_\text{ion}\rangle /$erg Hz$^{-1}$ $= 24.66_{-0.47}^{+0.18}$ for $M_\text{lim} = -11$ and $\log \langle f_\text{esc}\xi_\text{ion}\rangle /$erg Hz$^{-1}$ $= 25.02_{-0.46}^{+0.18}$ for $M_\text{lim} = -17$. However, the measurement of $\Gamma$ from the Lyman-$\alpha$ forest is far more model-dependent at $z=6$ than $z=5$ due to ongoing hydrogen reionisation. While the independent determination of \citet{Davies24} is consistent with \citet{Gaikwad23}, a model of reionisation (and/or a fluctuating UVB) is fundamentally required for the inference. Therefore, we show the measurements at $z=6$ as open points on Figure~\ref{fig1}. They are in fairly good agreement with measurements from individual galaxies and slightly above the LAE model from \citet{Matthee22} at $z\sim6$.

\section{Discussion}
\chang{We briefly consider the implications of our constraints on the average ionising escape fraction of galaxies at $z=5$. Direct measurements at $z\sim5$ have yielded population-averaged values of the emissivity of $\xi_\text{ion}\sim25.3$ (see e.g.~\citealt{Simmonds24}; \citealt{Stefanon22} and references therein). If this value holds for the entire galaxy population, our constraints from the Lyman-$\alpha$ forest require a population-averaged escape fraction of $f_\text{esc} = 9.5_{-3.5}^{+6.0}\%$. For a higher average $\xi_\text{ion}\sim25.5$, as has been reported by some studies (e.g.~\citealt{Bouwens16,Harikane18}), the escape fraction needs to be even lower in order to avoid making the forest too transmissive: $f_\text{esc}=6.0_{-2.2}^{+3.5}\%$. If the production of ionising photons is limited to bright LAEs with $M_\text{UV}<-17$ as in the model of \citet{Matthee22}, they would require an $f_\text{esc}=9.1_{-3.4}^{+5.7}\%$ assuming a $\xi_\text{ion}=25.5$. 
Later, at $z=4$ the global escape fraction is stringently required to be $f_\text{esc}<15\%$. Any contribution to the UV ionising background from active galactic nuclei at these redshifts would lower the required global galaxy escape fraction.}

\chang{This escape fraction of $\sim 10\%$ is similar to the initial predictions of the ionising photon budget required for star-forming galaxies to reionise the Universe \citep{Robertson13,Robertson15}. }

\chang{While galaxies observed by JWST at very early times do indeed show signs of elevated ionising photon production and/or escape, it is important to point out that reionisation must have a ``soft landing'' at $\lesssim5.3$. Theories of the ionising photon production of early galaxies (and quasars) need to be confronted to the effects they would have on the $z=5$ post-reionisation IGM -- where the total $\dot{N}_\text{ion}$ is already known robustly from observing the IGM itself.}

The argument in this paper is not new; it has appeared in a similar form in e.g.~\citet{Madau17,Shull12,Kuhlen12}. Theoretical models of the IGM have often implicitly incorporated the constraint on $\dot{N}_\text{ion}$ coming from measurements of $\Gamma$; see e.g.~\citet{Kakiichi18} and \citet{Davies21}. In contrast with these studies, we wished to directly present up-to-date constraints on the more familiar quantity $\langle f_\text{esc}\xi_\text{ion}\rangle$, which is of particular interest in light of ongoing observations of $z>5$ galaxies with the \textit{James Webb} Space Telescope (e.g.~\citealt{Saxena24,TT24}; Protu\v{s}ov\'{a} et al.~in prep). 


\begin{acknowledgements}

We thank the anonymous referee for their insightful comments which helped direct the discussion in the manuscript. 

The idea for this analysis came from productive discussions with Koki Kakiichi and others at the NORDITA workshop programme ``Cosmic Dawn at High Latitudes''. 
SEIB is supported by the Deutsche Forschungsgemeinschaft (DFG) under Emmy Noether grant number BO5771/1-1.

\end{acknowledgements}

\bibliographystyle{aa}
\bibliography{aanda}

\end{document}